\documentclass[12pt]{article}
\usepackage{amsfonts}

\newcommand{\Nr}{\mbox{\vphantom{$N$}\smash{$N/r$}}}
\newcommand{\Nt}{\mbox{\vphantom{$N$}\smash{$N/t$}}}

\DeclareMathSymbol{\square}{\mathord}{AMSa}{"03}
\newtheorem{theorem}{Theorem}
\newtheorem{corollary}[theorem]{Corollary}
\newenvironment{proof}{\begin{trivlist}\item[]{\flushleft\bf Proof }}
     {\hspace*{\fill}\raisebox{-1pt}{\boldmath$\square$}\end{trivlist}}

\title{Quantum Algorithm for the Collision Problem}

\author{
Gilles Brassard\,%
\thanks{\,Supported in part by Canada's {\sc nserc}, Qu\'ebec's {\sc
fcar}, and the Canada Council.}\\
{\protect\small\sl Universit\'e de Montr\'eal\/}\,%
\thanks{\,D\'epartement IRO, Universit\'e de Montr\'eal,
C.P. 6128, succursale centre-ville, Montr\'eal (Qu\'ebec), 
Canada H3C 3J7.  Email: \{\texttt{brassard},\texttt{tappa}\}%
\texttt{$\mathchar"40$iro.umontreal.ca}.}
\and
Peter H{\o}yer\,%
\thanks{\,Supported in part by the {\sc esprit} Long Term
Research Programme of the EU under project number 20244 ({\sc
alcom-it}).  
Research carried out while this author was at the Universit\'e de
Montr\'eal.}\\ {\protect\small\sl Odense University\/}\,%
\thanks{\,Department of Mathematics and Computer Science, Odense
University, Campusvej~55, \mbox{DK--5230} \mbox{Odense~M}, Denmark.
Email: \texttt{u2pi$\mathchar"40$imada.ou.dk}.}
\and
Alain Tapp\,%
\thanks{\,Supported in part by a postgraduate fellowship from Canada's
{\sc nserc}.}\\
{\protect\small\sl Universit\'e de Montr\'eal\/}\,\footnotemark[2]
}

\date{1~May 1997}

\begin{document}

\maketitle
\thispagestyle{empty}

\begin{abstract}
In this note, we give a quantum algorithm that finds collisions in
arbi\-trary \mbox{$r$-to-one} functions after only
\mbox{$O(\sqrt[3]{\Nr}\,)$}
expected evaluations of the function.
\mbox{Assuming} the function is given by a black box, this is
more efficient than the best possible classical algorithm,
even allowing probabilism.
We~also give a similar algorithm for finding claws in pairs of
functions.
Furthermore, we~exhibit a space-time tradeoff for our technique.
Our approach uses Grover's quantum searching algorithm in a novel~way.
\end{abstract}

\section{Introduction}
A {\em collision\/} for function~\mbox{$F:X \rightarrow Y$} consists of
two
distinct elements \mbox{$x_0,x_1 \in X$} such that
\mbox{$F(x_0)=F(x_1)$}. 
The~{\em collision problem\/} is to find a collision in~$F$ under the
promise that there is~one.

This problem is of particular interest for cryptology because some
functions known as {\em hash functions\/} are used in various
cryptographic protocols.  The secu\-rity of these protocols depends
crucially on the presumed difficulty of finding collisions in such
functions.
A~related question is to find so-called {\em claws\/} in pairs of
functions;
our quantum algorithm extends to this task.  This has consequences for
the
security of classical signature and bit commitment schemes.  We~refer
the
interested reader to~\cite{Stinson95} for general background on
cryptography,
which is not required for understanding our new collision-finding
algorithm.

A~function $F$ is said to be {\em $r$-to-one\/} if every element in its
image has exactly $r$ distinct preimages.
We~assume throughout this note that function~$F$ is given as a
black~box, so that it is not possible to obtain knowledge about it by
any
other means than evaluating it on points in its domain.
When $F$ is \mbox{two-to-one},
the most efficient classical algorithm possible for the collision
problem
requires an expected $\Theta(\sqrt{N}\,)$ evaluations of~$F$, where
\mbox{$N=|X|$} denotes the cardinality of the domain.  This classical
algorithm, which uses a principle reminiscent of the birthday paradox,
is reviewed in the next section. 

Recently, at a talk held at AT\&T, Eric Rains~\cite{Rains97} asked if it
is possible to do better on a quantum computer.  In~this note, we give a
positive answer to this question by providing a quantum algorithm that
finds
a collision in an arbitrary \mbox{two-to-one} function $F$ after only
$O(\sqrt[3]{N}\,)$ expected evaluations.

Earlier, Simon~\cite{Simon94} addressed the {\em {\sc xor}-mask
problem\/}
defined as follows.  Consider integers \mbox{$m \ge n$}.  We~are given a
function \mbox{$F: \{0,1\}^n \rightarrow \{0,1\}^m$} and promised that
either~$F$ is \mbox{one-to-one} or it is \mbox{two-to-one} and there
exists
an~\mbox{$s \in \{0,1\}^n$} such that \mbox{$F(x_0)=F(x_1)$}
if and only if \mbox{$x_0 \oplus x_1 = s$}, for all distinct
\mbox{$x_0, x_1 \in \{0,1\}^n$}, where~$\oplus$ denotes the bitwise
\mbox{exclusive-or}.  Simon's problem is to decide which of these two
conditions holds, and to find~$s$ in the latter case.
Note that finding $s$ is equivalent to finding a collision in the
case that $F$ is \mbox{two-to-one}.
Simon gave a quantum algorithm to solve his problem in
expected time polynomial in~$n$ and in the time required to compute~$F$. 
The~running time required for this task on a quantum computer was
recently
improved to being worst-case (rather than expected) polynomial time
thanks to
a more sophisticated algorithm~\cite{BH97}.  Simon's algorithm is
interesting
from a theoretical point of view because any classical algorithm that
uses only
sub-exponentially (in~$n$) many evaluations of~$F$ cannot hope to
distinguish
between the two types of functions significantly better than simply by
tossing
a coin, assuming equal {\em a~priori} probabilities~\cite{Simon94,BH97}. 
Unfortunately, the {\sc xor}-mask constraint when $F$ is
\mbox{two-to-one}
is so restrictive that Simon's algorithm has not yet found a practical
application. 

More recently, Grover~\cite{Grover96} discovered a quantum algorithm for
a
different searching problem.  We~are given a function
\mbox{$F:X \rightarrow \{0,1\}$} with the promise that there exists a
unique
\mbox{$x_0 \in X$} so that \mbox{$F(x_0)=1$}, and we are asked to
find~$x_0$. 
Provided the domain of the function is of cardinality a power of two
(\mbox{$N=2^n$}), Grover gave a quantum algorithm that finds the unknown
$x_0$ with probability at least~$1/2$ after only $\Theta(\sqrt{N}\,)$
evaluations of~$F$.

A~natural generalization of this searching problem occurs when
\mbox{$F:X \rightarrow Y$} is an arbitrary function.  Given some
\mbox{$y_0 \in Y$}, we are asked to find an \mbox{$x \in X$} such that
\mbox{$F(x)=y_0$}, provided such an~$x$ exists.  
If~\mbox{$t=|\{x \in X \, | \, F(x)=y_0\}|$} denotes the number of
different
solutions, \cite{BBHT96}~gives a generalization of Grover's algorithm
that
can find a solution whenever it exists (\mbox{$t \geq 1$}) after an
expected number of $O(\sqrt{\Nt}\,)$ evaluations of~$F$.
Although the algorithm does not need to know the value of $t$ ahead of
time,
it is more efficient (in~terms of the hidden constant in the $O$
notation)
when $t$ is known, which will be the case for most
algorithms given here.  From now on, we refer to this generalization of
Grover's algorithm as~$\mbox{\bf Grover}(F,y_0)$.
Note that the number of evaluations of~$F$ is not polynomially bounded
in~$\log N$ when~\mbox{$t \ll N$}; nevertheless Grover's algorithm is
considerably more efficient than classical brute-force searching.

In~the next section, we give our new quantum algorithm for solving the
collision problem for \mbox{two-to-one} functions.  We~then discuss a
straightforward generalization to \mbox{$r$-to-one} functions and even
to arbitrary functions whose image is sufficiently smaller than their
domain.  A~natural space-time tradeoff emerges for our technique.
Finally, we give applications to finding claws in pairs of functions.

\section{Algorithms for the collision problem}
We~first state two simple algorithms for the collision problem, one
classical and one quantum.  Both of these algorithms use an expected
number of $O(\sqrt{N}\,)$ evaluations of the given function, but the
quantum algorithm is more space efficient.
We~derive our improved algorithm from these two simple solutions.

The first solution is a well-known classical probabilistic
algorithm, here stated in slightly different terms than traditionally. 
The algorithm consists of three steps.  First, it selects a random
subset
\mbox{$K \subseteq X$} of cardinality \mbox{$k = c \sqrt{N}$} for an
appropriate constant~$c$.  Then, it computes the pair~\mbox{$(x,F(x))$}
for
each \mbox{$x \in K$} and sorts these pairs according to the second
entry. 
Finally, it outputs a collision in~$K$ if there is one, and otherwise
reports
that none has been found.  Based on the birthday paradox, it is not
difficult
to show that if~$F$ is \mbox{two-to-one} then this algorithm returns a
collision with probability at least~$1/2$ provided $c$ is sufficiently
large
(\mbox{$c \approx 1.18$} will~do).
If~we take a pair~\mbox{$(x,F(x))$} as unit of space then the algorithm
can be
implemented in space $\Theta(\sqrt{N}\,)$, and $\Theta(\sqrt{N}\,)$
evaluations of~$F$ suffice to succeed with probability~$1/2$.
If~we care about running time rather than simply the number of
evaluations
of~$F$, it may be preferable to resort to universal
hashing~\cite{CarterWegman}
rather than sorting to find a collision in~$K$.  This would avoid
spending
\mbox{$\Theta(\sqrt{N} \log N)$} time sorting the table, making possible
a
$\Theta(\sqrt{N}\,)$ overall expected running time if we assume that
each
evaluation of $F$ takes constant time.  We~stick to the sorting paradigm
for simplicity and because it is not clear if the benefits of universal
hashing carry over to quantum parallelism situations such as~ours.
We~come back to this issue in Section~\ref{sect:disc}.

The simple quantum algorithm for two-to-one functions also consists of
three
steps.  First, it picks an arbitrary element \mbox{$x_0 \in X$}.  Then,
it
computes \mbox{$x_1= \textbf{Grover}(H,1)$} where
\mbox{$H:X \rightarrow \{0,1\}$} denotes the function defined by
\mbox{$H(x)=1$} if and only if \mbox{$x \neq x_0$} and
\mbox{$F(x)=F(x_0)$}. 
Finally, it outputs the collision \mbox{$\{x_0,x_1\}$}.  There is
exactly one
\mbox{$x \in X$} that satisfies \mbox{$H(x)=1$} so~\mbox{$t=1$}, and
thus the
expected number of evaluations of~$F$ is also \mbox{$O(\sqrt{N}\,)$},
still to succeed with probability~$1/2$, but constant space suffices.

Our new algorithm, denoted {\bf Collision} and given below, can be
thought of as the logical union of the two algorithms above.  The main
idea is to select a subset~$K$ of~$X$ and then use {\bf Grover} to find
a collision \mbox{$\{x_0,x_1\}$} with \mbox{$x_0 \in K$} and
\mbox{$x_1 \in X \setminus K$}.  The~expected number of evaluations
of~$F$ and
the space used by the algorithm are determined by the parameter
\mbox{$k=|K|$}, the cardinality of~$K$.

\bigskip\noindent $\textbf{Collision}(F,k)$
\begin{enumerate}
\item \label{step:one}
      Pick an arbitrary subset \mbox{$K \subseteq X$} of
cardinality~$k$.
      Construct a table~$L$ of size~$k$ where each item in~$L$ holds a
      distinct pair \mbox{$(x,F(x))$} with \mbox{$x \in K$}.
\item Sort~$L$ according to the second entry in each item of~$L$.
\item \label{step:check} Check if $L$ contains a collision, that is,
      check if there exist distinct elements
      \mbox{$(x_0,F(x_0)), (x_1,F(x_1)) \in L$} for which
      \mbox{$F(x_0) = F(x_1)$}.
      If~so, goto step~\ref{step:last}.
\item \label{step:grover} Compute \mbox{$x_1 = \textbf{Grover}(H,1)$}
where
      \mbox{$H:X \rightarrow \{0,1\}$} denotes the function defined by
      \mbox{$H(x)=1$} if and only if there exists \mbox{$x_0 \in K$} so
that
      \mbox{$(x_0,F(x)) \in L$} but \mbox{$x \not= x_0$}.
      (Note that $x_0$ is unique if it exists since we already checked
      that there are no collisions in~$L$.)
\item Find \mbox{$(x_0,F(x_1)) \in L$}.
\item Output the collision \mbox{$\{x_0,x_1\}$}. \label{step:last}
\end{enumerate}

\begin{theorem}\label{thm:twotoone}
Given a two-to-one function \mbox{$F:X \rightarrow Y$} with
\mbox{$N=|X|$}
and an integer \mbox{$1 \leq k \leq N$}, algorithm
\mbox{$\textnormal{\bf Collision}(F,k)$} returns a collision after an
expected
number of \mbox{$O(k+\sqrt{N/k}\,)$} evaluations of~$F$ and uses
space~$\Theta(k)$. In~particular, when \mbox{$k=\sqrt[3]{N}$} then
\mbox{$\textnormal{\bf Collision}(F,k)$} evaluates~$F$ an expected
number of
\mbox{$O(\sqrt[3]{N}\,)$} times and uses space
\mbox{$\Theta(\sqrt[3]{N}\,)$}.
\end{theorem}

\begin{proof} 
The correctness of the algorithm follows easily from the definition
of~$H$ and the construction of $\textbf{Grover}(H,1)$.

We~now count the number of evaluations of~$F$.  In~the first step, the
algorithm uses~$k$ such evaluations.  Set $t=|\{x \in X \, | \,
H(x)=1\}|$.
By~the previous section, subroutine {\bf Grover} in
step~\ref{step:grover} uses an expected number of $O(\sqrt{\Nt}\,)$
evaluations of the function~$H$ to find one of the $t$ solutions.  Each
evaluation of $H$ can be implemented by using only one evaluation
of~$F$.  Finally, our algorithm evaluates~$F$ once in the penultimate
step,
giving a total expected number of $k+O(\sqrt{\Nt}\,)+1$ evaluations
of~$F$.
Since $F$ is two-to-one, $t$ equals the cardinality of~$K$, that is,
$t=k$, and the first part of the theorem follows.  The second part is
immediate.
\end{proof}

In~a nutshell, the improvement of our algorithm over
the simple quantum algorithm is achieved by trading time for space.
Suppose~the cardinality of set~$K$ is large.  Then the expected
number of evaluations of~$H$ used by subroutine
$\textbf{Grover}(H,1)$ is small, but on the other hand more space is
needed to store table~$L$.  Analogously, if~$K$ is small then the
space requirements are less but also $\textbf{Grover}(H,1)$ runs slower.

Suppose now that we apply algorithm {\bf Collision}, not necessarily on
a
\mbox{two-to-one} function, but on an arbitrary \mbox{$r$-to-one}
function
where \mbox{$r \geq 2$}.  Then we have the following theorem, whose
proof
is essentially the same as that of Theorem~\ref{thm:twotoone}.

\begin{theorem}\label{thm:rtoone}
Given an \mbox{$r$-to-one} function \mbox{$F:X \rightarrow Y$} with
\mbox{$r \geq 2$} and an integer \mbox{$1 \leq k \leq N=|X|$},
algorithm \mbox{$\textnormal{\bf Collision}(F,k)$} returns a collision
after
an expected number of \mbox{$O(k+\sqrt{N/(rk)}\,)$} evaluations of~$F$
and uses
space~$\Theta(k)$. In~particular, when \mbox{$k=\sqrt[3]{\Nr}$} then
\mbox{$\textnormal{\bf Collision}(F,k)$} uses an expected number of
\mbox{$O(\sqrt[3]{\Nr}\,)$} evaluations of~$F$ and space
\mbox{$\Theta(\sqrt[3]{\Nr}\,)$}.
\end{theorem}

Note that algorithm \mbox{$\textnormal{\bf Collision}(F,k)$}
can also be applied on an arbitrary function
\mbox{$F:X \rightarrow Y$} for which \mbox{$|X| \ge r |Y|$}
for some \mbox{$r > 1$}, even if $F$ is not \mbox{$r$-to-one}.
However, the algorithm must be modified in two ways for the general
case. 
First of all, the subset \mbox{$K \subseteq X$} of cardinality~$k$ must
be
picked at random, rather than arbitrarily, at step~\ref{step:one}.
Furthermore, the~fully generalized version of Grover's \mbox{algorithm}
given
in~\cite{BBHT96} must be used at step~\ref{step:grover} because the
number
of solutions for \mbox{$\textbf{Grover}(H,1)$} is no longer known in
advance
to be exactly \mbox{$t=(r-1)k$}.

By~varying~$k$ in Theorem~\ref{thm:rtoone}, the following space-time
tradeoff emerges.

\begin{corollary}
There exists a quantum algorithm that can find a collision in an
arbitrary
\mbox{$r$-to-one} function \mbox{$F:X \rightarrow Y$},
for any \mbox{$r \geq 2$}, using space $S$ and an expected number of
$O(T)$
evaluations of~$F$ for every $1 \leq S \leq T$ subject to
\[S T^2 \geq |F(X)| \]
where $F(X)$ denotes the image of~$F$.
\end{corollary}

Consider now two functions \mbox{$F:X \rightarrow Z$}
and \mbox{$G:Y \rightarrow Z$} that have the same codomain.
By~definition, a {\em claw\/} is a pair \mbox{$x \in X$},
\mbox{$y \in Y$} such that \mbox{$F(x)=G(y)$}.
Many cryptographic protocols are based on the assumption
that there are efficiently-computable functions $F$ and $G$
for which claws cannot be found efficiently even though
they exist in large number.

The simplest case arises when both $F$ and $G$ are bijections,
which is the usual situation when such functions are used to create
unconditionally-concealing bit commitment schemes~\cite{BCC88}.
If~\mbox{$N=|X|=|Y|=|Z|$}, algorithm {\bf Collision} is easily
modified as follows.

\bigskip\noindent $\textbf{Claw}(F,G,k)$
\begin{enumerate}
\item \label{claw:one}
      Pick an arbitrary subset \mbox{$K \subseteq X$} of
cardinality~$k$.
      Construct a table~$L$ of size~$k$ where each item in~$L$ holds a
      distinct pair \mbox{$(x,F(x))$} with \mbox{$x \in K$}.
\item Sort~$L$ according to the second entry in each item of~$L$.
\item \label{claw:grover} Compute \mbox{$y_0 = \textbf{Grover}(H,1)$}
where
      \mbox{$H:Y \rightarrow \{0,1\}$} denotes the function defined by
      \mbox{$H(y)=1$} if and only if a pair \mbox{$(x,G(y))$} appears
      in~$L$ for some arbitrary~\mbox{$x \in K$}.
\item Find \mbox{$(x_0,G(y_0)) \in L$}.
\item Output the claw \mbox{$(x_0,y_0)$}. \label{claw:last}
\end{enumerate}

\begin{theorem}\label{thm:clawtwotoone}
Given two one-to-one functions \mbox{$F:X \rightarrow Z$}
and \mbox{$G:Y \rightarrow Z$} with \mbox{$N=|X|=|Y|=|Z|$} and an
integer \mbox{$1 \leq k \leq N$}, algorithm $\textnormal{\bf
Claw}(F,G,k)$
\mbox{returns} a claw after $k$ evaluations of~$F$
and \mbox{$O(\sqrt{N/k}\,)$} evaluations of~$G$, and uses
space~$\Theta(k)$. In~particular, when \mbox{$k=\sqrt[3]{N}$} then
$\textnormal{\bf Claw}(F,G,k)$ evaluates~$F$ and $G$ an 
expected number of
\mbox{$O(\sqrt[3]{N}\,)$} times and uses space
\mbox{$\Theta(\sqrt[3]{N}\,)$}.
\end{theorem}

\begin{proof} 
Similar to the proof of Theorem~\ref{thm:twotoone}.
\end{proof}

The case in which both $F$ and $G$ are \mbox{$r$-to-one} for some
\mbox{$r \ge 2$} and \mbox{$N=|X|=|Y|=r|Z|$} is handled similarly.
However, it becomes necessary in step~\ref{claw:one} of algorithm
{\bf Claw} to select the elements of $K$ so that no two of them are
mapped to the same point by~$F$.  This will ensure that the call on
\mbox{$\textbf{Grover}(H,1)$} at step~\ref{claw:grover} has exactly
$kr$ solutions to choose from.  The~simplest way to choose~$K$ is
to pick random elements in~$X$ until \mbox{$|F(K)|=k$}.
As~long as \mbox{$k \le |Z|/2$}, this requires trying less than
$2k$ random elements of~$X$, except with vanishing probability.
The~proof of the following theorem is again essentially
as before.

\begin{theorem}\label{thm:clawrtoone}
Given two \mbox{$r$-to-one} functions \mbox{$F:X \rightarrow Z$}
and \mbox{$G:Y \rightarrow Z$} with \mbox{$N=|X|=|Y|=r|Z|$} and an
integer \mbox{$1 \leq k \leq N/2r$}, modified algorithm
$\textnormal{\bf Claw}(F,G,k)$ returns a claw after an 
expected number of
$\Theta(k)$ evaluations of~$F$ and
\mbox{$O(\sqrt{N/(rk)}\,)$} evaluations of~$G$, and uses
space~$\Theta(k)$.  In~particular, when \mbox{$k=\sqrt[3]{\Nr}$} 
then $\textnormal{\bf Claw}(F,G,k)$ evaluates~$F$ and $G$ an 
expected number of
\mbox{$O(\sqrt[3]{\Nr}\,)$} times and uses space
\mbox{$\Theta(\sqrt[3]{\Nr}\,)$}.
\end{theorem}

\section{Discussion}\label{sect:disc}
When we say that our quantum algorithms require $\Theta(k)$ space
to hold table~$L$, this corresponds unfortunately to the amount of
{\em quantum\/} memory, a~rather scarce resource with current
technology. 
Note however that this table is built classically in the initial 
steps of algorithms {\bf Collision} and {\bf Claw}: it needs to 
live in quantum memory for read purposes only.  
In~practice, it may be easier to build large read-only 
quantum memories than general read/write memories.

We considered only the number of evaluations of~$F$
in the analysis of algorithm {\bf Collision}.
The~time spent sorting~$L$ and doing
binary search in~$L$ should also be taken into account if we 
wanted to analyse
the running time of our algorithm.  If~we assume that it 
takes time~$T$ to compute the function (rather than assuming 
that it is given as a black box),
then it is straightforward to show that the algorithm given by
Theorem~\ref{thm:rtoone} runs in expected time
\[O((k+\sqrt{N/(kr)}\,) (T+\log k))\,.\]
Thus, the time spent sorting is negligible only if it takes
$\Omega(\log k)$ time to compute~$F$.
Similar considerations apply to algorithm {\bf Claw}.
It~is tempting to try using universal hashing to bypass the
need for sorting, as in the simple classical algorithm,
but it is not clear that this approach saves time here because
our use of quantum parallelism when we apply Grover's algorithm
will take a time that is given by the {\em maximum\/} time taken 
for all requests to the table, which is unlikely to be constant
even though the expected {\em average\/} time is constant.


\end{document}